Quantification of linear entropy for quantum entanglement in He, H- and Ps- ions using highly-correlated Hylleraas functions

Chien-Hao Lin[1,2], Yen-Chang Lin[1,3], and Yew Kam Ho[1]
[1]Institute of Atomic and Molecular Sciences, Academia Sinica, Taipei, Taiwan
[2]Department of Physics, National Taiwan Universities, Taiwan
[3]Graduate Institute of Applied Science and Engineering, Catholic Fu-Jen University, Taiwan

Abstract
The quantum entanglement for the two electrons in three-body atomic systems such as the helium atom, the hydrogen negative ion and the positronium negative ion are investigated by employing highly correlated Hylleraas functions to represent the ground states of such systems. As a measure of the spatial entanglement, the linear entropy of the reduced density matrix is calculated for the ground states. The required four-electron (12-dimensional) integrals are solved analytically such that they are suitable for machine computations. Results are compared with other calculations when available.

## 1. Introduction

There has been considerable interest in the investigations of quantum entanglement in atomic systems ([1-4], and references therein) including the two-electron model atoms, helium-like ions and helium atoms, as understanding of entangled systems is important in other research areas such as quantum information [5], quantum computation [6] and quantum cryptography [7]. The recent developments in entanglement for atomic and molecular systems were reviewed by Tichy *et al*. [8]. Works on model atoms such as the Moshinsky atom [9-12], the Crandall atom [1, 2, 13] and the Hooke atom [14, 15], have also been reported in the literatures, as well as the works on quantum dots [16-20]. In related developments, investigations on quantum entanglement of bosons in harmonic traps [21, 22] have also been carried out. For atomic systems, Osenda and Serra have calculated the von Neumann entropy in spherical helium [23]. Coe and D'Amico calculated the linear entropy for the ground state of the natural helium atom with wave functions constructed by using the products of hydrogenic wave functions, as well as using the density functional theory [2]. Manzano *et al*. [1] and Dehesa *et al*. [3, 4] investigated the entanglement of the helium ground and excited states using Kinoshita-type wave functions. These authors employed the Monte Carlo multidimensional integration scheme to treat the 12-dimensional integrals needed for calculations of linear entropy. Lin *et al* [24] employed configuration interaction with B-spline basis to represent the helium wave functions and calculated the linear entropy for the ground and excited states, and for the helium-like ions up to $Z=15$. Benenti *et al* [25] calculated the linear entropy and von Neumann entropy for the helium atom using configuration interaction basis wave functions constructed with Slater type orbitals (STO). In the present work, we report calculations of linear entropies for spatial entanglement (electron-electron orbital entanglement) in some three-particle atomic systems, such as the hydrogen negative ion, the positronium negative ion and the ground state helium atom by employing elaborate Hylleraas wave functions up to 203 terms. The energy-optimized wave functions are used to calculate the linear entropies. Our treatment of the four-electron integrals is different from the numerical integration used in Refs. [1, 3, 4]. By systematically changing the size of the expansion sets, we have established benchmark values for the linear entropy with relatively small estimated uncertainties in the He atom, and the H⁻ and Ps⁻ ions, with the measure of entanglement for the latter system is quantified for the first time in the literature. Atomic units are used in the present work.

## 2. Theoretical Method

The non-relativistic Hamiltonian (in atomic units) describing the three-body atomic systems are

$$H = -\frac{1}{2m_1}\nabla_1^2 - \frac{1}{2m_2}\nabla_2^2 - \frac{1}{2m_3}\nabla_3^2 - \frac{Z}{r_{13}} - \frac{Z}{r_{23}} + \frac{1}{r_{12}}, \qquad (1)$$

where 1, 2, and 3 denote the electron 1, electron 2 and the positively charged particle respectively, and $r_{ij}$ is the relative distance between the particle $i$ and $j$. When $m_3$ is infinitely heavy, the Hamiltonian with $Z=1$ and $Z=2$ are for the H⁻ ion and the He atom, respectively. When $Z=1$ and $m_3=1$ it is then for the positronium negative ion, Ps⁻. For $S$-states we use Hylleraas-type wave functions to describe the system, with

$$\Psi_{kmn}(\mathbf{r_{13}}, \mathbf{r_{23}}) = \sum_{kmn} C_{kmn} \left\{ \exp\left[-\alpha(r_{13} + r_{23})\right] r_{12}^k r_{13}^m r_{23}^n + (1 \leftrightarrow 2) \right\}, \qquad (2)$$

where $k+m+n \leq \omega$ and $\omega$, $l$, $m$ and $n$ are positive integers or zero. In the present work we use wave functions up to $N=203$ terms, with $\omega=11$.

The entanglement of atomic systems is quantified using different entanglement entropies, such as the von Neumann and linear entropies. In this work, we focus on the linear entropy. The von Neumann entropy has the form

$$S = -Tr(\rho \ln \rho), \qquad (3)$$

and the linear entropy is an approximation of the von Neumann entropy by taking

$$\ln x = (x-1) - \frac{(x-1)^2}{x} \cdots. \qquad (4)$$

By keeping the leading term in the above expansion, the linear entropy, denoted as $L_s$ becomes

$$L_s = -Tr(\rho(\rho-1)) = Tr(\rho - \rho^2) = 1 - Tr(\rho^2). \qquad (5)$$

The linear entropy for the two-electron systems is then defined as

$$L_s = 1 - Tr(\rho_{red}^2), \qquad (6)$$

where $\rho_{red}$ is the reduced density matrix. For two-component quantum systems, e.g., two-electron atoms, the reduced density matrix is obtained by tracing the two-particle density matrix over one of the two particles. To simplify the notation, from here on we denote $r_{13}$ by $r_1$ and $r_{23}$ by $r_2$. We have

$$\rho_{red}(\mathbf{r_1},\mathbf{r_2}) = \int \Psi^*(\mathbf{r_1},\mathbf{r_3}) \Psi(\mathbf{r_2},\mathbf{r_3}) d\mathbf{r_3}. \qquad (7)$$

The wave functions employed here are of Hylleraas type (see Eq.(2)). The advantage of using the correlation factors $r_{ij}$ explicitly enables us to take into account the correlation effects in an accurate and effective manner. But including the non-separable variable $r_{ij}$ leads to some complication as

$$Tr\rho_{red}^2 = \iint \rho_{red}(\mathbf{r}_1,\mathbf{r}_2)\rho_{red}(\mathbf{r}_2,\mathbf{r}_1)d\mathbf{r}_1 d\mathbf{r}_2$$
$$= \iint \left[\int \Psi^*(\mathbf{r}_1,\mathbf{r}_3)\Psi(\mathbf{r}_2,\mathbf{r}_3)d\mathbf{r}_3\right]\left[\int \Psi^*(\mathbf{r}_2,\mathbf{r}_4)\Psi(\mathbf{r}_1,\mathbf{r}_4)d\mathbf{r}_4\right]d\mathbf{r}_2 d\mathbf{r}_1 \quad (8)$$
$$= \iint\iint \Psi^*(\mathbf{r}_1,\mathbf{r}_3)\Psi(\mathbf{r}_2,\mathbf{r}_3)\Psi^*(\mathbf{r}_2,\mathbf{r}_4)\Psi(\mathbf{r}_1,\mathbf{r}_4)d\mathbf{r}_1 d\mathbf{r}_2 d\mathbf{r}_3 d\mathbf{r}_4.$$

This implies that some four-electron integrals are needed to be solved, and they have the general form of

$$I_4(i,j,k,l,m,n,p,q,s,t,a,b,c,d) = \int r_1^i r_2^j r_3^k r_4^l r_{12}^m r_{13}^n r_{14}^p r_{23}^q r_{24}^s r_{34}^t e^{-ar_1-br_2-cr_3-dr_4} d\mathbf{r}_1 d\mathbf{r}_2 d\mathbf{r}_3 d\mathbf{r}_4. \quad (9)$$

In the Appendix of this paper, we present the detailed results of Eq. (8) in terms of the $I_4$ integrals. In Refs.[1, 3, 4], the authors solved this kind of four-electron integrals by carrying out the 12-dimensional integrals using Monte Carlo multidimensional numerical integration routines. But using such an approach their results may have statistical uncertainty for the calculated linear entropy, not so small for some excited states, and that their results may even lead to errors in numerical calculations [4]. In the present work, we prefer to carry out the four-electron integrals analytically to forms that are suitable for machine computations. There have been considerable references for four-electron integrals in the literatures (for example, see Refs. [26, 27], and references therein), and we have found the seminal work by King [26] suit our purpose quite well. We have re-derived the four-electron integral (9) following [26, 27], and have obtained the identical results as those in Ref [26]. (The typo errors in [26] were corrected in [27].) We have then independently written up the code for computer computation. To check our code, we have obtained the same numerical results for different input parameters for the $I_4$ integrals as those reported in Table I of Ref. [26]. Together with the needed routines for two- and three-electron integrals [28 – 31], we are able to calculate the linear entropies for the helium atom, the hydrogen negative ion, and the positronium negative ion by employing Hylleraas functions up to 203 terms ($\omega$=11). Finally, we should mention that in the present work we only consider the spatial (electron-electron orbital) entanglement for these three-body atomic systems. As for the entanglement coming from the spin part of the two electrons (that form a singlet-spin state), readers are referred to the well-discussed earlier publications [1-3, 13, 14, 20, 25], and we will not repeat them here.

## 3. Results and Discussion

We have calculated the linear entropy for the H$^-$ ion, Ps$^-$ ion and the ground state He atom using wave functions of Eq. (2). Different expansion lengths are used, with $\omega$ =6, 7, 8, 9, 10, 11, corresponding to number of terms $N$ = 50, 70, 95, 125, 161, 203, respectively. The energy in each term is individually optimized, and the wave functions are then used to calculate the linear entropy $L_s$ using Eq. (6) – (8), and Eq. (13) in Appendix. Results are shown in Tables 1, 2, 3, for He, H$^-$ and Ps$^-$, respectively. Furthermore, we can achieve better results by extrapolating $\omega$ to infinity. Since the even and odd powers of the $r_{12}$ factors contribute to the energy differently, we extrapolate the energy and linear entropy according to even or odd number of $\omega$ values separately, i.e, $\omega$ =(6, 8, 10) or $\omega$ =(7, 9, 11). Using a geometric fit to represent the convergence, we can extrapolate to infinite for the energy ($E$) as follows (see [32], for example):

$$\Delta(\omega) \equiv E(\omega) - E(\omega-2), \quad (10)$$

$$E(\infty) = E(\omega) + \frac{\Delta^2(\omega)}{\Delta(\omega-2) - \Delta(\omega)}. \quad (11)$$

In Tables 1 to 3, we show comparisons for our energies with those of elaborate calculations in the literature [27 – 29]. A formula similar to the above can also be applied to extrapolate linear entropy [$E$ is replaced by $L_s$ in Eq. (10) and (11)] as well. The final results representing benchmark values for linear entropy in two-electron atomic systems are shown in Tables 1, 2, and 3 for the ground state of the helium atom, the hydrogen negative ion, and the positronium negative ion, respectively. For the ground state helium atom, we conclude $L_s = 0.0159156 \pm 0.0000010$, and a comparison with earlier calculations [3, 24, 25] is also made in Table 1. Our present result represents an improvement over previous values in the literature as it has a smaller estimated uncertainty in our calculation. For the hydrogen negative ion, we determine our result as $L_s = 0.106153 \pm 0.000010$. Again, with smaller estimated uncertainty, we believe our present value is also an improvement over other calculations, notwithstanding that no explicit numerical value was reported in Ref. [1]. Finally, we determine the linear entropy for the positronium negative ion as $L_s = 0.120796 \pm 0.000010$. To the best of our knowledge, there is no reported linear entropy for Ps⁻ in the literature for comparison. In addition to the investigation of quantum entanglement for H⁻ and Ps⁻ ions, we have also calculated the linear entropy for the model systems between these two natural ions with different masses of the positively charged particle. Figure 1 shows the linear entropy vs $1/m$, with m being the mass (in units of electron mass) of the positively charged particle. It is seen that the linear entropy increases from the hydrogen negative ion to the positronium negative ion for increasing $1/m$ (decreasing mass $m$). This indicates that as the mass of the positively charged particle decreases, the two electrons become more correlated, resulting in the increase of spectral entanglement for these three-body ions from H⁻ to Ps⁻. Figure 2 shows the linear entropy vs the nuclear charge $Z$ from H⁻ ($Z=1$) to He ($Z=2$). It is seen that with decreasing $Z$ from helium to H⁻, the correlation effect between the two electrons would increase, leading to the increase of the linear entropy, in a manner consistent with previous findings [24] that the linear entropy was found increasing when the nuclear charge of the helium sequence was decreased from $Z=15$ to $Z=2$.

## 4. Summary and Conclusion

We have investigated the spatial entanglement of the ground state helium atom, the hydrogen negative ion and the positronium negative ion using elaborate Hylleraas basis wave functions. The linear entropies for these systems are calculated with the 12-dimensional integrals being solved analytically to expressions that are suitable for machine computations. Convergences of the linear entropies in terms of expansion lengths in the wave functions are examined, establishing benchmark values with relative small estimated uncertainties for these three-body atomic systems that are useful references for future investigations on quantum entanglement in natural atomic systems.

Acknowledgements
The present work is supported by National Science Council of Taiwan. C.-H. Lin is a recipient of Dr. Chau-Ting Chang Scholarship, and is supported by NSC with grant no. 102-2815-C-001-004-M.

Appendix: Evaluation of $Tr\rho_{red}^2$

In equation (8), the trace of $\rho_{red}^2$ is represented as an integral over the position vectors of four electrons, where each wavefunction $\Psi$ is of Hylleraas type,

$$\Psi_{kmn}(\mathbf{r_1},\mathbf{r_2}) = \sum_{kmn} C_{kmn} \left\{ \exp\left[-\alpha(r_1+r_2)\right] r_1^k r_2^m r_{12}^n + (1 \leftrightarrow 2) \right\}. \tag{12}$$

The linear entropy of entanglement $L_s = 1 - Tr\rho_{red}^2$, and $Tr\rho_{red}^2$ can be expanded in terms of integrals for the Hylleraas basis:

$$\begin{aligned}
Tr\rho_{red}^2 = & \sum_{k_1 m_1 n_1} \sum_{k_2 m_2 n_2} \sum_{k_3 m_3 n_3} \sum_{k_4 m_4 n_4} C_{k_1 m_1 n_1} C_{k_2 m_2 n_2} C_{k_3 m_3 n_3} C_{k_4 m_4 n_4} \iint\iint e^{-2\alpha(r_1+r_2+r_3+r_4)} \\
& \times \{ r_1^{k_1+k_4} r_{13}^{n_1} r_{14}^{n_4} r_2^{k_2+k_3} r_{23}^{n_2} r_{24}^{n_3} r_3^{m_1+m_2} r_4^{m_3+m_4} + r_1^{k_1+k_4} r_{13}^{n_1} r_{14}^{n_4} r_2^{k_2+m_3} r_{23}^{n_2} r_{24}^{n_3} r_3^{m_1+m_2} r_4^{k_3+m_4} \\
& + r_1^{k_1+k_4} r_{13}^{n_1} r_{14}^{n_4} r_2^{m_2+k_3} r_{23}^{n_2} r_{24}^{n_3} r_3^{m_1+k_2} r_4^{m_3+m_4} + r_1^{k_1+k_4} r_{13}^{n_1} r_{14}^{n_4} r_2^{m_2+m_3} r_{23}^{n_2} r_{24}^{n_3} r_3^{m_1+k_2} r_4^{k_3+m_4} \\
& + r_1^{k_1+m_4} r_{13}^{n_1} r_{14}^{n_4} r_2^{k_2+k_3} r_{23}^{n_2} r_{24}^{n_3} r_3^{m_1+m_2} r_4^{m_3+k_4} + r_1^{k_1+m_4} r_{13}^{n_1} r_{14}^{n_4} r_2^{k_2+m_3} r_{23}^{n_2} r_{24}^{n_3} r_3^{m_1+m_2} r_4^{k_3+k_4} \\
& + r_1^{k_1+m_4} r_{13}^{n_1} r_{14}^{n_4} r_2^{m_2+k_3} r_{23}^{n_2} r_{24}^{n_3} r_3^{m_1+k_2} r_4^{m_3+k_4} + r_1^{k_1+m_4} r_{13}^{n_1} r_{14}^{n_4} r_2^{m_2+m_3} r_{23}^{n_2} r_{24}^{n_3} r_3^{m_1+k_2} r_4^{k_3+k_4} \\
& + r_1^{m_1+k_4} r_{13}^{n_1} r_{14}^{n_4} r_2^{k_2+k_3} r_{23}^{n_2} r_{24}^{n_3} r_3^{k_1+m_2} r_4^{m_3+m_4} + r_1^{m_1+k_4} r_{13}^{n_1} r_{14}^{n_4} r_2^{k_2+m_3} r_{23}^{n_2} r_{24}^{n_3} r_3^{k_1+m_2} r_4^{k_3+m_4} \\
& + r_1^{m_1+k_4} r_{13}^{n_1} r_{14}^{n_4} r_2^{m_2+k_3} r_{23}^{n_2} r_{24}^{n_3} r_3^{k_1+k_2} r_4^{m_3+m_4} + r_1^{m_1+k_4} r_{13}^{n_1} r_{14}^{n_4} r_2^{m_2+m_3} r_{23}^{n_2} r_{24}^{n_3} r_3^{k_1+k_2} r_4^{k_3+m_4} \\
& + r_1^{m_1+m_4} r_{13}^{n_1} r_{14}^{n_4} r_2^{k_2+k_3} r_{23}^{n_2} r_{24}^{n_3} r_3^{k_1+m_2} r_4^{m_3+k_4} + r_1^{m_1+m_4} r_{13}^{n_1} r_{14}^{n_4} r_2^{k_2+m_3} r_{23}^{n_2} r_{24}^{n_3} r_3^{k_1+m_2} r_4^{k_3+k_4} \\
& + r_1^{m_1+m_4} r_{13}^{n_1} r_{14}^{n_4} r_2^{m_2+k_3} r_{23}^{n_2} r_{24}^{n_3} r_3^{k_1+k_2} r_4^{m_3+k_4} + r_1^{m_1+m_4} r_{13}^{n_1} r_{14}^{n_4} r_2^{m_2+m_3} r_{23}^{n_2} r_{24}^{n_3} r_3^{k_1+k_2} r_4^{k_3+k_4} \} \\
& \times d\mathbf{r}_1 d\mathbf{r}_2 d\mathbf{r}_3 d\mathbf{r}_4.
\end{aligned} \tag{13}$$

Each part of the sum is in a general form of $I_4$ integral (see Eq.(9)), and the solution for the $I_4$ integral is given in Refs. [26, 27].

**References**

1. Manzano, D., Plastino, A.~R., Dehesa, J.~S., and Koga, T.: Quantum entanglement in two-electron atomic models, *J. Phys. A : Math. Theor.* **43**, 275301 (2010).
2. Coe, J.~P. and D'Amico, I.: The entanglement of few-particle systems when using the local-density approximation, *J. Phys. : Conf. Ser.* **254**, 012010 (2010).
3. Dehesa, J. S., Koga, T., Yáñez, R. J., Plastino, A. R., and Esquivel, R. O.: Quantum entanglement in helium, *J. Phys. B: At. Mol. and Opt. Phys.* **45**, 015504 (2012).
4. Dehesa, J. S., Koga, T., Yáñez, R. J., Plastino, A. R., and Esquivel, R. O.: Corrigendum: Quantum entanglement in helium, *J. Phys. B: At. Mol. and Opt. Phys.* **45**, 239501 (2012).
5. Nielsen, M. A. and Chuang, I. L.: *Quantum Computation and Quantum Information*, Cambridge university press (2010).
6. Zhao, M.-J., Fei, S.-M., and Li-Jost, X.: Complete entanglement witness for quantum teleportation, *Phys. Rev. A* **85**, 054301 (2012).
7. Naik, D. S., Peterson, C. G., White, A. G., Berglund, A. J., and Kwiat, P. G.: Entangled state quantum cryptography: Eavesdropping on the Eckert protocol, *Phys. Rev. Lett.* **84**, 4733 (2000).
8. Tichy, M. C., Mintert, F., and Buchleitner, A.: Essential entanglement for atomic and molecular physics, *J. Phys. B: At. Mol. and Opt. Phys.* **44**, 192001 (2011).
9. Moshinsky, M.: How good is the Hartree-Fock approximation, *Am. J. Phys.* **36**, 52 (1968).
10. Amovilli, C. and March, N. H.: Exact density matrix for a two-electron model atom and approximate proposals for realistic two-electron systems, *Phys. Rev. A* **67**, 022509 (2003).
11. Amovilli, C. and March, N. H.: Quantum information: Jaynes and Shannon entropies in a two-electron entangled artificial atom, *Phys. Rev. A* **69**, 054302 (2004).
12. Nagy, I. and Pipek, J.: Approximations for the interparticle interaction energy in an exactly solvable two-electron model atom, *Phys. Rev. A* **81**, 014501 (2010).
13. Yaüez, R. J., Plastino, A. R., and Dehesa, J. S.: Quantum entanglement in a soluble two-electron model atom, *Euro. Phys. J. D* **56**, 141 (2010).
14. Coe, J. P., Sudbery, A., and D'amico, I.: Entanglement and density-functional theory: Testing approximations on hooke's atom, *Phys. Rev. B* **77**, 205122 (2008).
15. Koscik, P. and Hassanabadi, H.: Entanglement in hooke's law atoms: an effect of the dimensionality of the space, *Few-Body Systems* **52**, 189 (2012).
16. Abdullah, S., Coe, J. P., and D'Amico, I.: Effect of confinement potential geometry on entanglement in quantum dot-based nanostructures, *Phys. Rev. B* **80**, 235302 (2009).
17. Nazmitdinov, R. G., Simonovic, N. S., Plastino, A. R., and Chizhov, A. V.: Shape transitions in excited states of two-electron quantum dots in a magnetic field, *J. Phys. B: At. Mol. and Opt. Phys.* **45**, 205503 (2012).
18. Okopinska, A. and Koscik, P.: Correlation and entanglement in elliptically deformed two-electron quantum dots, *Few-Body Systems* **50**, 413 (2011).
19. Coden, D. S. A., Romero, R. H., Ferrón, A., and Gomez, S.~S.: Impurity effects in two-electron coupled quantum dots: entanglement modulation, *J. Phys. B: At. Mol. and Opt. Phys.* **46**, 065501 (2013).

20. Schröter, S., Friedrich, H., and Madroñero, J.: Considerations on Hund's first rule in a planar two-electron quantum dot, *Phys. Rev. A* **87**, 042507 (2013).
21. Koscik, P.: Quantum correlations of a few bosons within a harmonic trap, *Few-Body Systems* **52**, 49 (2012).
22. Okopinska, A. and Koscik, P.: Entanglement of two charged bosons in strongly anisotropic traps, *Few-Body Systems* **54**, 629 (2013).
23. Osenda, O. and Serra, P.: Scaling of the von Neumann entropy in a two-electron system near the ionization threshold, *Phys. Rev. A* **75**, 042331 (2007).
24. Lin, Y.-C., Lin, C.-Y., and Ho, Y. K.: Spatial entanglement in two-electron atomic systems, *Phys. Rev. A* **87**, 022316 (2013).
25. Benenti, G., Siccardi, S., and Strini, G.: Entanglement in helium, *Euro. Phys. J. D* **67**, 1 (2013).
26. King, F. W.: Analysis of some integrals arising in the atomic four-electron problem, *J. Chem. Phys.* **99**, 3622 (1993).
27. King, F. W.: Reply to "comment on `analysis of some integrals arising in the atomic four-electron problem`" [*J. Chem Phys.* **99**, 3622(1993)], *J. Chem. Phys.* **120**, 3042 (2004).
28. Ho, Y. K. and Page, B. A. P.: Evaluation of some integrals required in low-energy electron or positron-atom scattering, *J. Comput. Phys.*, **17**, 122 (1975).
29. Perkins, J. F.: Evaluation of a four-electron atomic integral, *J. Comput. Phys.* **17**, 434 (1975).
30. Sims, J. S. and Hagstrom, S. A.: One-center $r_{ij}$ integrals over Slater-type orbitals, *J. Chem. Phys.* **55**, 4699 (1971).
31. Ohrn, Y. and Nordling, J.: On the calculation of some atomic integrals containing functions of $r_{12}$, $r_{13}$, and $r_{23}$, *J. Chem. Phys.* **39**, 1864 (1963).
32. Drachman, R. J., Ho, Y. K., and Houston, S. K.: Positron attachment to helium in the $^3S$ state, *J. Phys. B: At. Mol. and Opt. Phys.* **9**, L199 (1976).
33. Drake, G. W. F.: High precision calculations for helium, *Atomic, Molecular, and Optical Physics Handbook*, pp 154--171 (1996).
34. Drake, G. W. F., Cassar, M. M., and Nistor, R. A.: Ground-state energies for helium, H$^-$, and Ps$^-$, *Phys. Rev. A* **65**, 054501 (2002).
35. Ho, Y. K.: Variational calculation of ground-state energy of positronium negative ions, *Phys. Rev. A* **48**, 4780 (1993).

Table I. The energy and linear entropy of the helium atom in is ground state.

| ω | Number of terms | Energy (atomic unit) | Linear Entropy |
|---|---|---|---|
| 3 | 13 | -2.903640446 | 0.015886739 |
| 4 | 22 | -2.903713944 | 0.015922114 |
| 5 | 34 | -2.903720967 | 0.015915605 |
| 6 | 50 | -2.903723702 | 0.015916146 |
| 7 | 70 | -2.903724105 | 0.015915692 |
| 8 | 95 | -2.903724305 | 0.015915709 |
| 9 | 125 | -2.903724344 | 0.015915656 |
| 10 | 161 | -2.903724366 | 0.015915657 |
| 11 | 203 | -2.903724372 | 0.015915649 |
| Extrapolated (6, 8, 10) | | -2.903724373 | 0.015915650 |
| Extrapolated (7, 9, 11) | | -2.903724375 | 0.015915648 |
| Present final estimate | | | 0.0159156 ± 0.0000010 |
| Drake and coworkers [33, 34] | | -2.9037243770341196 | |
| Y. C. Lin *et al* [24] | | -2.9035869 | 0.015943±0.00004 |
| Dehesa *et al* [3] | | -2.903724377 | 0.015914±0.000044 |
| Benenti *et. al*.[25] | | | 0.01606 |

Table II, The energy and linear entropy of the hydrogen negative ion.

| $\omega$ | Wave Functions | Energy (atomic unit) | Linear Entropy |
|---|---|---|---|
| 5 | 34 | -0.527707183 | 0.10553029 |
| 6 | 50 | -0.527743248 | 0.10605049 |
| 7 | 70 | -0.527747857 | 0.10610703 |
| 8 | 95 | -0.527750064 | 0.10613936 |
| 9 | 125 | -0.527750614 | 0.10614786 |
| 10 | 161 | -0.527750865 | 0.10615100 |
| 11 | 203 | -0.527750943 | 0.10615256 |
| Extrapolated (6, 8, 10) | | -0.527750972 | 0.10615276 |
| Extrapolated (7, 9, 11) | | -0.527750988 | 0.10615317 |
| Present final estimate | | | 0.106153±0.000010 |
| Drake and coworkers [32, 33] | | -0.527751016544377196503 | |

Table III. The energy and linear entropy of the positronium negative ion

| ω | Number of terms | Energy (atomic unit) | Linear Entropy |
|---|---|---|---|
| 5 | 34 | -0.261957583 | 0.119612504 |
| 6 | 50 | -0.262001051 | 0.120694682 |
| 7 | 70 | -0.262002801 | 0.120723743 |
| 8 | 95 | -0.262004669 | 0.120786171 |
| 9 | 125 | -0.262004894 | 0.120791288 |
| 10 | 161 | -0.262005012 | 0.120794606 |
| 11 | 203 | -0.262005044 | 0.120796128 |
| Extrapolated (6,8,10) | | -0.262005048 | 0.12079546 |
| Extrapolated (7,9,11) | | -0.262005056 | 0.12079650 |
| Present final estimate | | | 0.120796±0.000010 |
| Drake and coworkers [34] | | -0.262005070232980107627 | |
| Ho [35] | | -0.26200507023294 | |

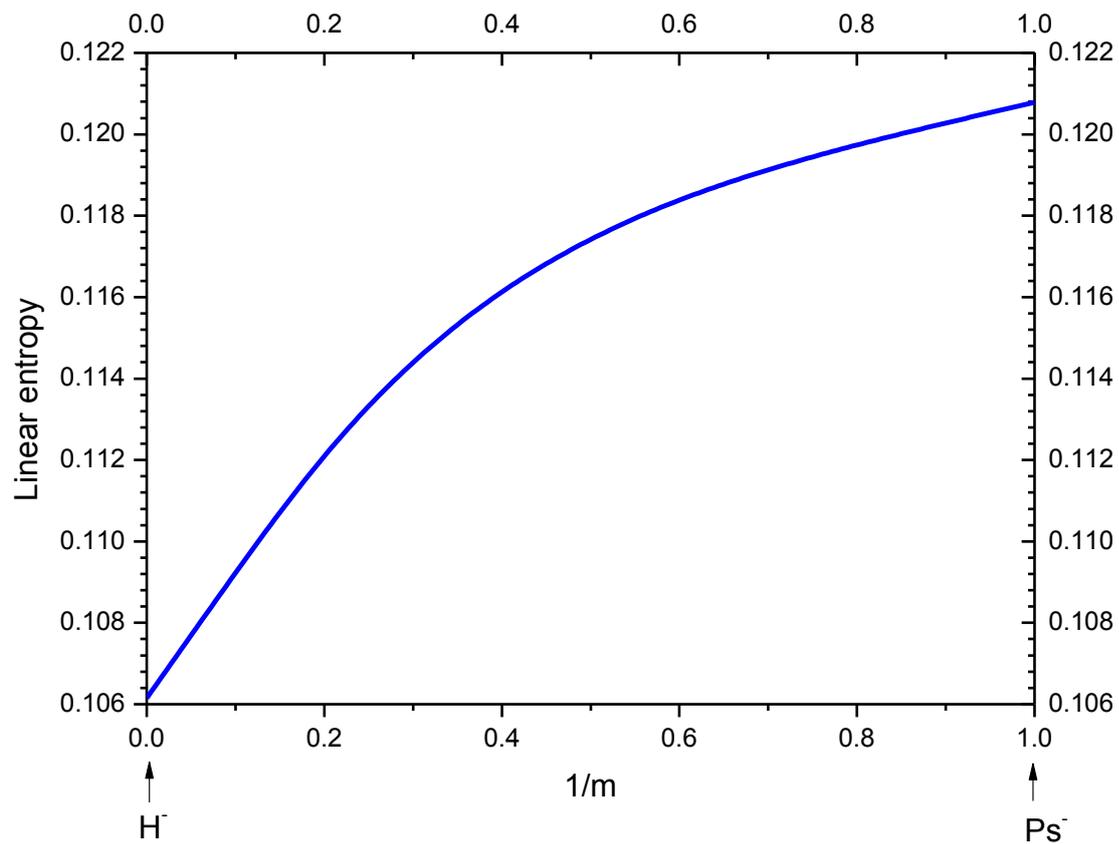

Figure 1. Linear entropy for hydrogen negative ion to the positronium negative ion for changing $1/m$, with $m$ being the mass (in units of electron mass) of the positively charged particle.

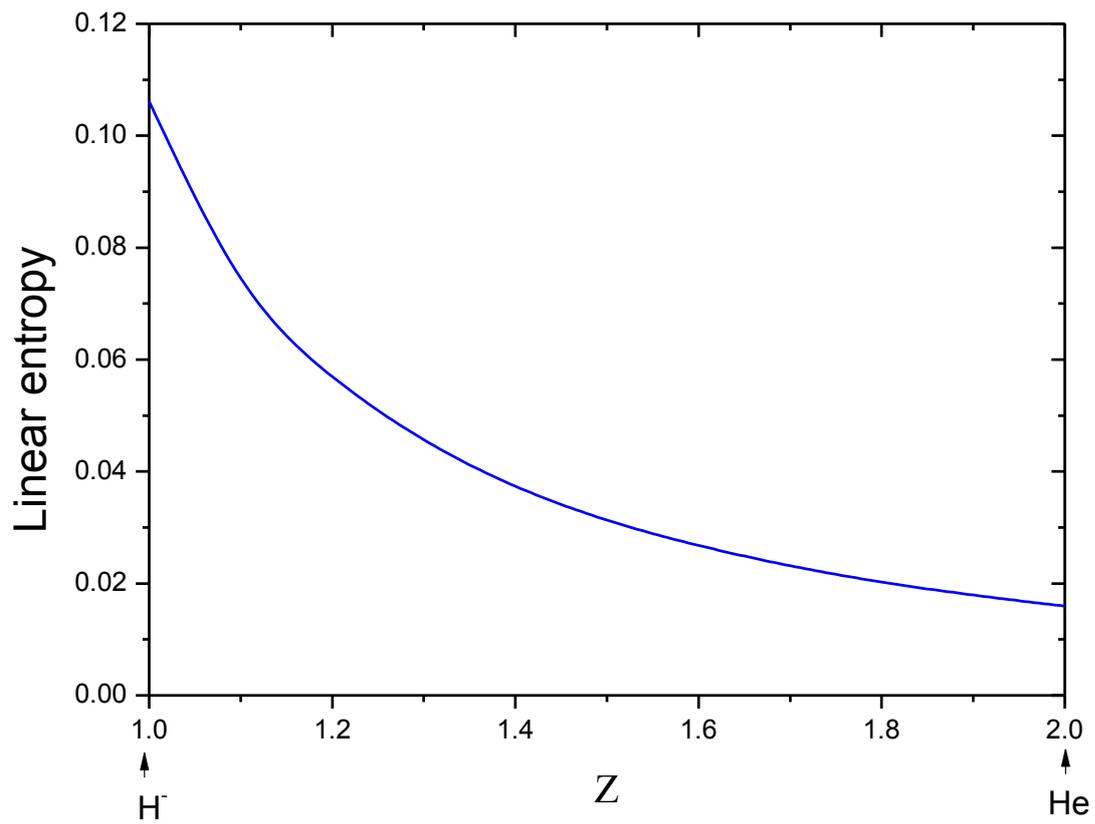

Figure 2, Linear entropy for the hydrogen negative ion ($Z=1$) to the helium atom ($Z=2$) for changing $Z$, with $Z$ being the charge of the positively charged particle.